\documentclass[11pt,a4paper]{article}
\pdfoutput=1
\usepackage{jheppub}
\usepackage{graphics}
\usepackage{stmaryrd}
\usepackage{amsfonts}
\usepackage{mathrsfs}
\usepackage{color}
\usepackage{bm}
\usepackage{amsmath}
\usepackage{chemarrow}
\def\dbar{{\mathchar '26\mkern -10mu\delta}}

\def\be{\begin{equation}}
\def\ee{\end{equation}}
\def\ba{\begin{eqnarray}}
\def\ea{\end{eqnarray}}

\def\nn{\nonumber}

\newcommand{\eq}[1]{(\ref{#1})}

\def\dbar{{\mathchar '26\mkern -10mu\delta}}

\def\q{\theta} \def\w{\omega}\def\r {\rho} \def\t {\tau} \def\y {\psi}   \def\p {\pi} \def\a {\alpha} \def\s {\sigma} \def\d {\delta} \def\f {\phi} \def\g {\gamma} \def\h {\eta} \def\j {\varphi} \def\k {\kappa} \def\l {\lambda} \def\z {\zeta} \def\x {\xi} \def\c {\chi} \def\b {\beta}  \def\m {\mu} \def\pd {\partial}\def\p {\pi} \def \inf {\infty}  
\def\Q{\Theta} \def\W{\Omega}     \def\S {\Sigma}  \def\F {\Phi}      \def\grad{\nabla}\def\.{\cdot}
\def\math {\mathcal}
\title{First Law of Black Hole in the Gravitational Electromagnetic System}

\author{Jie Jiang$^{ab}$}
\author{,Aofei Sang$^{ab}$}
\author{,and Ming Zhang$^{c}$\footnote{Corresponding author.}}

\affiliation[a]{College of Education for the Future, Beijing Normal University, Zhuhai 519087, China}

\affiliation[b]{Department of Physics, Beijing Normal University, Beijing, 100875, China}
\affiliation[c]{Department of Physics, Jiangxi Normal University, Nanchang 330022, China}

\emailAdd{jiejiang@mail.bnu.edu.cn, 202021140021@mail.bnu.edu.cn, mingzhang@jxnu.edu.cn}

\abstract{After considering the quantum corrections of Einstein-Maxwell theory, the effective theory will contain some higher-curvature terms and nonminimally coupled electromagnetic fields. In this paper, we study the first law of black holes in the gravitational electromagnetic system with the Lagrangian $\math{L}(g_{ab}, R_{abcd}, F_{ab})$. Firstly, we calculate the Noether charge and the variational identity in this theory, and then generically derive the first law of thermodynamics for an asymptotically flat stationary axisymmetrical symmetric black hole without the requirement that the electromagnetic field is smooth on the bifurcation surface. Our results indicate that the first law of black hole thermodynamics might be valid for the Einstein-Maxwell theory with some quantum corrections in the effective region.
}

\keywords{First law, stationary black hole, electromagnetic field}

\begin{document}

\maketitle

\section{Introduction}

General relativity is the most successful theory to describe the interaction of gravity. It predicts the existence of the black hole, which is a fundamental object in theoretical physics, astronomy, and cosmology.
Over the past few decades, many studies of general relativity have shown that black holes can be viewed as a thermodynamic system and satisfy the four laws of thermodynamics \cite{A2, A3, A31}. By considering the semi-classical quantum effect in curved spacetime, Hawking found that the black hole can be regarded as a blackbody system \cite{A4}, which provides a natural provided a natural explanation to the laws of black hole thermodynamics. After that, the thermodynamics of black holes has aroused wide interest among researchers, and people believe that it can give us a deeper understanding of gravity.

The most profound laws of black hole mechanics are the first and second laws. With a straightforward derivation, the first law of the Kerr-Newmann black hole shows the relationship between the variations of black hole mass $M$, angular momentum $J$, electric charge $Q$, and areas, i.e.,
\ba\label{firstlaw001}
\d M=\frac{\k}{8\p}\d A+\W_H\d J+\F_H\d Q\,,
\ea
in which $\k$, $\W_H$, and $\F_H$ are the surface gravity, angular velocity, and electric potential of the event horizon. The original derivation of the first law demands that the perturbation of the spacetime is stationary (``equilibrium state version'') \cite{A31}. Moreover, their calculation is also based on the Einstein equation. After that, the discussion is extended to the ``physical process version'', where a stationary black hole is changed by some infinitesimal physical process \cite{Sudarsky:1992ty,Iyer:1994ys}. In particular, Iyer and Wald \cite{Iyer:1994ys} show that the above first law of thermodynamic relation is also applicable to any diffeomorphism covariant theories, in which the first law of black holes can be regarded as a straightforward result of the variational identity, and it can be expressed as
\ba
\d M=\frac{\k}{2\p}\d S+\W_H\d J\,,
\ea
where
\ba
S\equiv -2\p \int_{\math{B}}\tilde{\bm{\epsilon}}\frac{\d \math{L}}{\d R_{abcd}}\hat{\bm{\epsilon}}_{ab}\hat{\bm{\epsilon}}_{cd}
\ea
is the Wald entropy, in which $\bm \epsilon_{ab}$ is the binormal of the cross-section $\math{B}$ of the event horizon. However, it is worth noting that the ``potential-charge'' term does not appear explicitly in the result derived by Iyer and Wald \cite{Iyer:1994ys}. The different result is caused by the assumption that the asymptotically flat stationary black hole contains a bifurcated Killing horizon and all fields are smooth on the Killing horizon as well as the bifurcation surface. In the gravitational electromagnetic system, because of the gauge covariance of the electromagnetic field, the vector potential is not a real physical quantity in the spacetime, and therefore it is not necessary to demand that it is smooth on the Killing horizon. With this consideration, Gao derived the first law of the asymptotically flat stationary black holes in Einstein-Maxwell and Einstein-Yang-Mills theories without the assumption that the vector potential is smooth on the Killing horizon \cite{Gao:2003ys}. Their result shows the same expression as Eq. \eq{firstlaw001}.

The standard Einstein-Maxwell is a good approximation to describe the gravitational and electromagnetic interactions at a low-energy regime. However, at higher-energy regime, the effective theory should be corrected by adding some higher-order derivative terms to take into account the quantum effects \cite{A13,A14,A15,A16}, including the higher-curvature terms and nonminimally coupled electromagnetic field terms. These corrections will modify the dynamics of gravity as well as the laws of black holes. A natural question is whether the first law of black holes is also satisfied in these effective theories after the quantum corrections are taken into account. Therefore, in this paper, we would like to extend Gao's discussion \cite{Gao:2003ys} into a more general gravitational electromagnetic system and derive the first law of black holes without the assumption that the vector potential is smooth at the Killing horizon.

The remainder of this paper is organized as follows. In the next section, we derive the explicit expressions of the Noether charge and variational identity in the gravitational electromagnetic theory with a general Lagrangian $\math{L}(g_{ab}, R_{abcd}, F_{ab})$. In section \ref{sec3}, after assuming that the metric and electromagnetic strength are smooth near the Killing horizon as well as the bifurcation surface, we derive the first law of the black hole thermodynamics in the gravitational electromagnetic system. Finally, we give a brief conclusion in section \ref{sec4}.

\section{Noether charge in the gravitational electromagnetic system}\label{sec2}
In this section, we first review the Noether current and Noether charge in the diffeomorphism covariant gravitational electromagnetic theory. The Lagrangian $n$-form is given by
\ba\begin{aligned}\label{Lg2}
\bm{L}=\bm{\epsilon} \math{L}(g_{ab}, R_{abcd}, F_{ab})\,,
\end{aligned}\ea
in which $\bm{F}=d\bm{A}$ with the vector potential $\bm{A}$ is the electromagnetic strength, $R_{abcd}$ is the Riemann curvature tensor of the Lorentz signature metric $g_{ab}$, and $\math{L}$ is an analytic function of the scalars from the contraction of $R_{abcd}$ and $F_{ab}$. In the following, we refer to $(g_{ab}, \y)$ as $\f$ collectively. Consider a one-parameter family $\f(\l)$ of the configuration space. The variation of any quantity $\h(\l)$ is defined by
\ba
\d \f=\left.\frac{d\f(\l)}{d\l}\right|_{\l=0}\,.
\ea
Variation of the Lagrangian $n$-form can be formally divided as
\ba\label{deltaL}
\d \bm{L}=\bm{E}_\f\d\f+d\bm{\Q}(\f, \d\f)\,,
\ea
in which $\bm{E}_\f=0$ is the equation of motion and $\bm{\Q}(\f, \d\f)$ is the symplectic potential of this theory. Next, we are going to calculate the explicit expression of these quantities. From Eq. \eq{Lg2}, we have
\ba\begin{aligned}
\d\bm{L}&=\bm{\epsilon}\d\math{L}+(\d\bm{\epsilon})\math{L}\\
&=\bm{\epsilon}\d\math{L}+\frac{1}{2}\bm{L}g^{ab}\d g_{ab}\,.
\end{aligned}\ea
For the first term of the above equation, we have
\ba\begin{aligned}\label{dL}
\d\math{L}&=A^{ab}\d g_{ab}+E_R^{abcd}\d R_{abcd}+E_F^{ab}\d F_{ab}\,,
\end{aligned}\ea
in which we have denoted
\ba\begin{aligned}
A^{ab}&=\frac{\pd \math{L}}{\pd g_{ab}}\,,\quad E_R^{abcd}=\frac{\pd \math{L}}{\pd R_{abcd}}\,,\quad E_F^{ab}=\frac{\pd \math{L}}{\pd F_{ab}}\,.
\end{aligned}\ea
For the first term of Eq. \eq{dL}, using the relation
\ba\begin{aligned}
A^{ab}\d g_{ab}=-\frac{\pd \math{L}}{\pd g^{ab}}\d g^{ab}\,,
\end{aligned}\ea
we have
\ba\begin{aligned}
A_{ab}=-\frac{\pd \math{L}}{\pd g^{ab}}\,.
\end{aligned}\ea
Considering the assumption that $\math{L}$ is a function of the contractions of $R_{abcd}$ and $F_{ab}$, it is not hard to get
\ba\begin{aligned}
\frac{\pd \math{L}}{\pd g^{ab}}&=2(E_R)_{a}{}^{cde} R_{bcde}+(E_F)_{a}{}^cF_{bc}\,.
\end{aligned}\ea
After noting that the index of ``$a$'' and ``$b$'' in the above expression is symmetric, we also have
\ba\begin{aligned}\label{symmetriceom}
2(E_R)_{[a}{}^{cde} R_{b]cde}+(E_F)_{[a}{}^cF_{b]c}=0\,.
\end{aligned}\ea
For the second term of Eq. \eq{dL}, we have
\ba\begin{aligned}
&E_R^{abcd}\d R_{abcd}=E_R^{abcd}R_{abc}{}^e\d g_{de}-2E_R^{acbd}\grad_d\grad_c\d g_{ab}\\
&=(E_R^{cdea}R_{cde}{}^b-2\grad_c\grad_dE_R^{acbd})\d g_{ab}+2\grad_d(\grad_c E_R^{adbc}\d g_{ab}-E_R^{acbd}\grad_c \d g_{ab}),
\end{aligned}\ea
For the third term of Eq. \eq{dL}, we have
\ba\begin{aligned}
E_F^{ab}\d F_{ab}=2E_F^{ab}\grad_{a}\d A_b
=-2\grad_aE_F^{ab}\d A_b+2\grad_d (E_F^{db}\d A_b)\,.
\end{aligned}\ea
Summing the above results, we can get
\ba\begin{aligned}
\d\math{L}=&-\left(E_R^{cdea}R_{cde}{}^b+2\grad_c\grad_dE_R^{acbd}+E_F^{ac}F^b{}_c\right)\d g_{ab}-2\grad_aE_F^{ab}\d A_b+\grad_d\dbar v^d\\
\end{aligned}\ea
with
\ba\begin{aligned}\label{dbarv}
\dbar v^d&=2\grad_c E_R^{adbc}\d g_{ab}-2 E_R^{acbd}\grad_c \d g_{ab}+2 E_F^{db}\d A_b\,.
\end{aligned}\ea
Using
\ba\begin{aligned}
\grad_d \dbar v^d=d\star \bm{\dbar v}\,,
\end{aligned}\ea
we can further obtain
\ba\begin{aligned}\label{ETheta}
\bm{\Q}(\f, \d\f)&=\bm{\Q}^\text{grav}(\f, \d g)+\bm{\Q}^\text{e.m.}(\f,\d\bm{A})\,,
\end{aligned}\ea
in which
\ba\begin{aligned}
\bm{\Q}^\text{grav}_{a_2\cdots a_n}(\f, \d g)&=\bm{\epsilon}_{ca_2\cdots a_n}\left(2\grad_dE_R^{acbd}\d g_{ab}+2 E_R^{abcd}\grad_b \d g_{ad}\right)\,,\\
\bm{\Q}^\text{e.m.}_{a_2\cdots a_n}(\f, \d \bm{A})&=2\bm{\epsilon}_{aa_2\cdots a_n}E_F^{ab}\d A_b\,.
\end{aligned}\ea
Moreover, we also have
\ba\begin{aligned}\label{ETheta}
\bm{E}_\f\d\f&=-\bm{\epsilon}\left(\frac{1}{2}T^{ab}\d g_{ab}+j^a\d A_a\right)\,,
\end{aligned}\ea
in which
\ba\begin{aligned}\label{eomchap02}
T^{ab}&=2E_R^{cde(a}R_{cde}{}^{b)}+4\grad_c\grad_dE_R^{(a|c|b)d}+2E_F^{(a|c|}F^{b)}{}_c-g^{ab}\math{L}\,,\\ j^b&=2\grad_a E_F^{ab}
\end{aligned}\ea
can be regarded as the stress-energy tensor and electric current of the extra matter source.

Using the symplectic potential, the symplectic current $(n-1)$-form is defined by
\ba\begin{aligned}
\bm{\w}(\f, \d_1\f, \d_2\f)=\d_2\bm{\Q}(\f, \d_1\f)-\d_1\bm{\Q}(\f, \d_2\f)\,,
\end{aligned}\ea
in which $\d_1$ and $\d_2$ are the variations related to any two different one-parameter families. If the spacetime $M$ is global hyperbolic, denoting $C$ to the Cauchy surface, the symplectic form of this theory is defined as
\ba\begin{aligned}
\W(\f, \d_1\f, \d_2\f)=\int_{C}\bm{\w}(\f, \d_1\f, \d_2\f)\,.
\end{aligned}\ea

The Noether current $(n-1)$-form related to the vector field $\z^a$ is defined as
\ba\begin{aligned}
\bm{J}_\z=\bm{\Q}(\f, \math{L}_\z\f)-\z\.\bm{L}\,.
\end{aligned}\ea
Using Eq. \eq{deltaL}, it is not hard to verify
\ba
d \bm{J}_\z=-\bm{E}_\f \math{L}_\z\f\,.
\ea
Therefore, if the dynamical field $\f$ satisfy the on-shell condition $\bm{E}_\f=0$, the Noether current is a closed form, i.e., $d\bm{J}_\z=0$, which implies there is a Noether charge $(n-2)$-form $\bm{Q}_\z$ such that $\bm{J}=d\bm{Q}$. Next, we prove the following lemma:\\
\textbf{Lemma 1} For the theory with Lagrangian \eq{Lg2}, the Noether current $\bm{J}_\z$ can be divided as
\ba\begin{aligned}\label{Jdef}
\bm{J}_\z=\bm{C}_\z+d\bm{Q}_\z\,,
\end{aligned}\ea
in which
\ba\begin{aligned}
(\bm{C}_\z)_{a_2\cdots a_n}&=\bm{\epsilon}_{a a_2\cdots a_n}(\z^bT_b{}^a+\z^bA_b j^a)\,,\\
(\bm{Q}_\z)_{a_3\cdots a_n}&=\bm{\epsilon}_{aba_3\cdots a_n}\left(E_F^{ab}A^c\z_c-2\grad_dE_R^{abcd}\z_c-E_R^{abcd}\grad_{[c}\z_{d]}\right)\,.
\end{aligned}\ea
are the constraint $(n-1)$-form and Noether charge $(n-2)$-form of this theory separately. When the dynamical field $\f$ satisfies the on-shell condition, we have $\bm{C}_\z=0$.\\
\textit{Proof.} The Noether current $\bm{J}_\z$ can be written as
\ba\begin{aligned}\label{Jexp}
\bm{J}_\z&=\bm{\Q}(\f, \math{L}_\z\f)-\z\.\bm{L}\\
&=\star (\bm{v}_\z-\z \math{L})\,,
\end{aligned}\ea
in which we have defined $\bm{v}_\z=\left.\dbar \bm{v}\right|_{\d\f=\math{L}_\z\f}$. Based on Eq. \eq{dbarv}, the first term of above equation can be expressed as
\ba\begin{aligned}
v_\z^c=&2E_R^{abcd}\grad_b(\math{L}_\z g_{ad})+2\math{L}_\z g_{bd} W^{cbd}+E_1^c\math{L}_\z\c\\
&+E_2^{cb}\grad_b (\math{L}_\z\c)-\grad_b E_2^{bc}\math{L}_\z\c+2 E_F^{cb}\math{L}_\z A_b\,.
\end{aligned}\ea
Using
\ba\begin{aligned}
\math{L}_\z g_{ab}&=2\grad_{(a} \z_{b)}\,,\\
\math{L}_\z A_a&=\grad_a(\z^b A_b)+\z^b F_{ba}\,,
\end{aligned}\ea
we have
\ba\begin{aligned}\label{vc}
v_\z^c&=4E_R^{abcd}\grad_b\grad_{(a}\z_{d)}+4\grad_{(b}\z_{d)} \grad_a E_R^{abcd}+2 E_F^{cb}\grad_b(\z^aA_a)+2E_F^{cb}\z^a F_{ab}\,.
\end{aligned}\ea
For the first term of above expression, we have
\ba\begin{aligned}
v_{12}^c=&2E_R^{abcd}\grad_b\grad_{a}\z_{d}+2E_R^{abcd}\grad_b\grad_{d}\z_{a}
+2\grad_{b}\z_{d} \grad_a E_R^{abcd}+2\grad_{d}\z_{b} \grad_aE_R^{abcd}\\
=&2E_R^{abcd}\grad_{[b}\grad_{a]}\z_{d}+4E_R^{abcd}\grad_{[b}\grad_{d]}\z_{a}
+2E_R^{abcd}\grad_d\grad_b\z_{a}\\
&+2\grad_{b}\z_{d} \grad_a E_R^{abcd}+2\grad_{d}\z_{b} \grad_aE_R^{abcd}\\
=&E_R^{abdc}R_{abde}\z^e+2E_R^{abcd}R_{bdae}\z^{e}+2\grad_d(E_R^{abcd}\grad_b\z_{a})\\
&-2\grad_b\z_{a}\grad_dE_R^{abcd}
+2\grad_{b}\z_{d} \grad_a E_R^{abcd}+2\grad_{d}\z_{b} \grad_aE_R^{abcd}\\
=&E_R^{abdc}R_{abde}\z^e+2E_R^{abcd}R_{bdae}\z^{e}+2\grad_d(E_R^{abcd}\grad_b\z_{a})-2\grad_b(\z_{a}\grad_dE_R^{abcd})\\
&+2\grad_{b}(\z_{d} \grad_a E_R^{abcd})+2\grad_{d}(\z_{b} \grad_aE_R^{abcd})+2\z_{a}\grad_b\grad_dE_R^{abcd}\\
&-2\z_{d}\grad_{b}\grad_aE_R^{abcd}-2\z_{b}\grad_{d}\grad_aE_R^{abcd}\\
=&E_R^{abdc}R_{abde}\z^e+2E_R^{abcd}R_{bdae}\z^{e}+2\z_d\grad_a\grad_b(E_R^{dacb}+ E_R^{abcd}-E_R^{bdca})\\
&+2\grad_d[E_R^{abcd}\grad_b\z_{a}+\z_{b}\grad_a(E_R^{abcd}-E_R^{dacb}-E_R^{bdca})]\,.
\end{aligned}\ea
From the definition of $E_R^{abcd}$, we can see that $E_R^{abcd}$ has the same symmetries as $R_{abcd}$. Therefore, $E_R^{abcd}$ also satisfies the Bianchi identity $E_R^{[abc]d}=0$, i.e.,
\ba\begin{aligned}
E_R^{abcd}+E_R^{bdca}+E_R^{dacb}=0\,.
\end{aligned}\ea
Using the above identity, we have
\ba\begin{aligned}
v_{12}^c=&E_R^{abdc}R_{abde}\z^e+2E_R^{abcd}R_{bdae}\z^{e}-4\z_d\grad_a\grad_bE_R^{bdca}\\
&+2\grad_d[E_R^{abcd}\grad_b\z_{a}+2\z_{b} \grad_aE_R^{abcd}]\\
=&E_R^{abdc}R_{abde}\z^e+2E_R^{abcd}R_{bdae}\z^{e}+4\z_d\grad_{(a}\grad_{b)}E_R^{cadb}+4\z_d\grad_{[a}\grad_{b]}E_R^{cadb}\\
&+2\grad_d[E_R^{abcd}\grad_b\z_{a}+2\z_{b} \grad_aE_R^{abcd}]\\
=&E_R^{abdc}R_{abde}\z^e+4\z_d\grad_{a}\grad_{b}E_R^{(c|a|d)b}+2E_R^{abcd}R_{bdae}\z^{e}-2R_{abe}{}^cE_R^{eadb}\z_d\\
&-2R_{abe}{}^d E_R^{caeb}\z_d+2\grad_d[E_R^{abcd}\grad_b\z_{a}+2\z_{b} \grad_aE_R^{abcd}].\\
\end{aligned}\ea
Considering the following results
\ba\begin{aligned}
2E_R^{abcd}R_{bdae}\z^{e}&=E_R^{abcd}R_{bdae}\z^{e}+E_R^{abcd}R_{dabe}\z^{e}
=E_R^{abdc}R_{abde}\z^{e}\,,\\
2R_{abe}{}^cE_R^{eadb}\z_d&=R_{abe}{}^cE_R^{abed}\z_d,\quad 2R_{abe}{}^dE_R^{caeb}\z_d=-R_{bea}{}^dE_R^{beac}\z_d,
\end{aligned}\ea
we can further obtain
\ba\begin{aligned}
v_{12}^c&=3E_R^{abdc}R_{abde}\z^e-E_R^{abde}R_{abd}{}^c\z_e
+4\z_d\grad_{a}\grad_{b}E_R^{(c|a|d)b}\\
&\quad+2\grad_d(E_R^{abcd}\grad_b\z_{a}+2\z_{b} \grad_aE_R^{abcd})\\
&=4E_R^{abd[c}R_{abd}{}^{e]}\z_e+2E_R^{abd(c}R_{abd}{}^{e)}\z_e+4\z_d\grad_{a}\grad_{b}E_R^{(c|a|d)b}\\
&\quad+2\grad_d(E_R^{abcd}\grad_b\z_{a}+2\z_{b} \grad_aE_R^{abcd})\,.
\end{aligned}\ea
For the third and forth terms of Eq. \eq{vc}, we have
\ba\begin{aligned}
v_{34}^c=2 E_F^{cb}\grad_b(\z^aA_a)+ 2E_F^{cb}\z^a F_{ab}=2\grad_d(E_F^{cd}\z^aA_a)+j^c A_a\z^a+ 2E_F^{cb}\z^a F_{ab}.
\end{aligned}\ea
Summing the above results, we have
\ba\begin{aligned}
v_{\z}^c&=3E_R^{abdc}R_{abde}\z^e-E_R^{abde}R_{abd}{}^c\z_e
+ 2E_F^{cb}\z^a F_{ab}+4\z_d\grad_{a}\grad_{b}E_R^{(c|a|d)b}\\
&\quad+j^c A_e\z^e+2\grad_d(E_R^{abcd}\grad_b\z_{a}+2\z_{b} \grad_aE_R^{abcd}+E_F^{cd}\z^aA_a)\\
&=(4E_R^{abd[c}R_{abd}{}^{e]}+2E_F^{b[c} F_{b}{}^{e]})\z_e+(2E_R^{abd(c}R_{abd}{}^{e)}+2E_F^{b(c} F^{e)}{}_{b}+4\grad_{a}\grad_{b}E_R^{(c|a|e)b})\z_e\\
&\quad+j^c A_e\z^e+2\grad_d(E_R^{abcd}\grad_b\z_{a}+2\z_{b} \grad_aE_R^{abcd}+E_F^{cd}\z^aA_a)\,.
\end{aligned}\nn\ea
From Eq. \eq{symmetriceom}, we can see that the first term of the above expression vanishes. Together with the equation of motion \eq{eomchap02}, we can get
\ba\begin{aligned}
v_\z^c-\z^c\math{L}=&\z^eT_e{}^c+\z^eA_e j^c+2\grad_d(E_R^{abcd}\grad_b\z_{a}+2\z_{b} \grad_aE_R^{abcd}+E_F^{cd}\z^aA_a)\,.
\end{aligned}\ea
Therefore, we have
\ba\begin{aligned}
\bm{J}_\z=\bm{C}_\z+d\bm{Q}_\z
\end{aligned}\ea
with
\ba\begin{aligned}
(\bm{C}_\z)_{a_2\cdots a_n}&=\bm{\epsilon}_{a a_2\cdots a_n}(\z^bT_b{}^a+\z^bA_b j^a)\,,\\
(\bm{Q}_\z)_{a_3\cdots a_n}&=\bm{\epsilon}_{aba_3\cdots a_n}\left(E_F^{ab}A^c\z_c-2\grad_dE_R^{abcd}\z_c-E_R^{abcd}\grad_{[c}\z_{d]}\right)\,.
\end{aligned}\ea
As we desired to show. $\Box$
\\

Variation of Noether current $\bm{J}_\z$ from Eq. \eq{Jdef}, we can get
\ba\begin{aligned}
\bar{\d} \bm{J}_\z&=\bar{\d} \bm{\Q}(\f, \math{L}_\z\f)-\z\.\d\bm{L}\\
&=\bar{\d} \bm{\Q}(\f, \math{L}_\z\f)-\z\.\bm{E}_\f\d\f-\z\.d\bm{\Q}(\f, \d\f)\\
&=\bar{\d} \bm{\Q}(\f, \math{L}_\z\f)-\math{L}_\z\bm{\Q}(\f, \d\f)+d[\z\.\bm{\Q}(\f, \d\f)]-\z\.\bm{E}_\f\d\f\\
&=\bm{\w}(\f, \d\f, \math{L}_\z\f)+d[\z\.\bm{\Q}(\f, \d\f)]-\z\.\bm{E}_\f\d\f\,,\\
\end{aligned}\ea
where we have introduce the notation $\bar{\d}$ to denote the variation when the vector field $\z^a$ is fixed, i.e., we have
\ba
\d X_\z=\bar{\d} X_{\z}+X_{\d\z}
\ea
for the quantity $X_{\z}$.

Moreover, using Eq. \eq{Jexp}, we have
\ba\begin{aligned}
\bar{\d}\bm{J}_\z=\bar{\d}\bm{C}_\z+d\bar{\d}\bm{Q}_\z\,.
\end{aligned}\ea
Combining the above results, we can obtain the following identity
\ba\begin{aligned}\label{id1chap01}
d[\bar{\d}\bm{Q}_\z-\z\.\bm{\Q}(\f, \d\f)]=\bm{\w}(\f, \d\f, \math{L}_\z\f)-\z\.\bm{E}_\f\d\f-\bar{\d}\bm{C}_\z\,.
\end{aligned}\ea

In the following, we consider a one-parameter family $\f(\l)$ in which any $\f(\l)$ satisfy the on-shell condition, i.e., we have $\bm{C}(\l)=\bm{E} _\f(\l)=0$ and $\d \bm{C}=\bm{E}_\f=0$. Then, the variational identity becomes
\ba\begin{aligned}\label{id2chap01}
\bm{\w}(\f, \d\f, \math{L}_\z\f)=d[\bar{\d}\bm{Q}_\z-\z\.\bm{\Q}(\f, \d\f)]\,.
\end{aligned}\ea

Consider the asymptotically flat stationary axisymmetric spacetime satisfying the asymptotic condition of ``Case I'' in Ref. \cite{Wald:1999wa}. Let $\z^a$ be a vector field related to the symmetry at asymptotically infinity. Then, there exists a conserved quantity $H_ {\z}$ related to this vector field. If we assume $\f(\l)$ satisfies the on-shell condition, $\d H_{\z}$ can be expressed as\cite{Wald:1999wa,Gao:2003ys}
\ba\begin{aligned}
\d H_{\z}=\int_{\inf}\left(\bar{\d}\bm{Q}_{\z}-\z\.\bm{\Q}\right)\,,
\end{aligned}\ea
in which ``$\inf$'' denotes a $(n-2)$-sphere at asymptotically infinity.
When $\z^a$ is chosen as the vector field $t^a$ related to the  asymptotic time translation or $\j^a$ related to the rotation, the canonical mass and angular momentum can be defined by\cite{Gao:2003ys}
\ba\begin{aligned}\label{MJcp2}
\d M&=\int_{\inf}\left(\bar{\d}\bm{Q}_{t}-t\.\bm{\Q}\right)\,,\quad\quad
\d J=\int_{\inf}\left(\bar{\d}\bm{Q}_{\j}-\j\.\bm{\Q}\right)\,.
\end{aligned}\ea
Using the equation of motion \eq{eomchap02}, the electric charge of the spacetime is defined by
\ba\begin{aligned}
Q&=-\int_{\inf} \bm{\epsilon}_{aba_3\cdots a_n} E_F^{ab}\,.
\end{aligned}\ea

\section{The first law of the stationary axisymmetric black holes}\label{sec3}

In this section, we would like to derive the first law of black holes in the gravitational electromagnetic system with Lagrangian \eq{Lg2}. Let $(M, g_{ab})$ is an asymptotically flat stationary axisymmetric spacetime satisfying the asymptotic condition of ``Case I'' in Ref. \cite{Wald:1999wa}, and there is a bifurcated Killing horizon $H$ with a bifurcated surface $\math{B}$. Assume that the metric $g_{ab}$ and electromagnetic strength $F_{ab}$ is smooth near the horizon as well as the bifurcation surface. The generated Killing vector field of the Killing horizon can be expressed as
\ba\begin{aligned}\label{xtwj}
\x^a=t^a+\W_H \j^a\,,
\end{aligned}\ea
in which we have denoted $\W_H \j^a=\W_H^{(\m)}\j^a_{(\m)}$. Here $t^a$ and $\j_{(\m)}^a$ are the Killing vector fields related to the time transition and axial symmetries of the spacetime, $\W_H^{(\m)}$ is the velocity of the black hole horizon $\math{H}$. In the following, we consider a one-parameter family $\f(\l)$, any element in which is a stationary axisymmetric black hole as described above. Considering the diffeomorphism invariance of the theory, we can choose a gauge such that $\x^a$ and Killing horizon $\math{H}$ (including the bifurcation surface $\math{B}$) is independent on $\l$, i.e., they are fixed under the variation. Replacing $\z^a$ by $\x^a$ and considering the symmetries
\ba
\math{L}_\x g_{ab}(\l)=0\,,\quad \math{L}_\x\bm{A}(\l)=0\,,
\ea
the variational identity \eq{id2chap01} implies
\ba\begin{aligned}\label{varid2}
d[\bar{\d} \bm{Q}_\x-\x\.\bm{\Q}(\f,\d \f)]=0\,.
\end{aligned}\ea
Choose $\S$ to a hypersurface connecting the sphere $S_\inf$ at infinity and a cross-section $S$ on the future Killing horizon. Integration of Eq. \eq{varid2} on $\S$, using the Stokes theorem, we can further obtain
\ba\begin{aligned}\label{first1}
\int_\math{\inf} \left[\bar{\d} \bm{Q}_\x-\x\.\bm{\Q}(\f,\d \f)\right]=\int_S \left[\bar{\d} \bm{Q}_\x-\x\.\bm{\Q}(\f,\d \f)\right]\,.
\end{aligned}\ea
For the left side of the above expression, using the definition of the mass and angular motion \eq{MJcp2}, we have
\ba\begin{aligned}
\int_\math{\inf} \left[\bar{\d} \bm{Q}_\x-\x\.\bm{\Q}(\f,\d \f)\right]&=\int_\math{\inf} \left[\bar{\d} \bm{Q}_\x-t\.\bm{\Q}(\f,\d \f)\right]\\
&=\int_\math{\inf} \left[\bar{\d} \bm{Q}_t-t\.\bm{\Q}(\f,\d \f)\right]+\int_\math{\inf} \bar{\d} \bm{Q}_{\W_H\j}\\
&=\d M-\W_H\d J\,.
\end{aligned}\ea
For the right side of Eq. \eq{first1}, considering the gauge choice $\d\x^a=0$, we can replace $\bar{\d}$ by $\d$. Then, we have
\ba\begin{aligned}\label{first2}
\int_S \left[\bar{\d} \bm{Q}_\x-\x\.\bm{\Q}(\f,\d \f)\right]=&\d\int_S\bm{\epsilon}_{aba_3\cdots a_n}\left[E_F^{ab}A^c\x_c-2\x_c \grad_d E_R^{abcd}-E_R^{abcd}\grad_{[c}\x_{d]}\right]\\
&-\int_S\x\.\left[\bm{\Theta}^\text{e.m}(\f,\d\bm{A})+\bm{\Theta}^\text{grav}(\f,\d g)\right] \,.
\end{aligned}\ea

Since we assume that $R_{abcd}$, $g_{ab}$ and $F_{ab}$ is smooth near the Killing horizon (including the bifurcation surface $\math{B}$),  $E_R^{abcd}$ and $E_F^{ab}$ would also be the smooth tensor near the horizon. Before deriving the first law, we first prove the following Lemma:

\noindent \textbf{Lemma 2} For a stationary black hole with bifurcated Killing horizon $\math{H}$. Let $\x^a$ be a Killing vector field generated the future Killing horizon, $S$ be a cross-section of future horizon, and $s^a$ is another null vector field on $\math{H}$ satisfying
\ba\begin{aligned}\label{sxws}
s^a \x_a=1\,,\quad s^a s_a=0\,, \quad w^a_is_a=0\,,
\end{aligned}\ea
in which $w_i^a$ is the tangent vector on the cross section $S$. Denote $z_i^a=\{\x^a, s^a, w^a_j\}$ to the basis on the cross section. Then, for any tensor field $X_{a_1\cdots a_k}$ which is smooth on the horizon (including bifurcation surface) and satisfies $\math{L}_\x X=X_{a_1\cdots a_k}$, if $X_{a_1\cdots a_k} z_1^a \cdots z_k^a$ is not zero, the number of $\x^a$ must not be greater than the number of $s^a$.

\noindent \textit{Proof.} Consider a foliation of the horizon $\math{H}$ which is obtained from the cross-section $S$ by the diffeomorphism generated by the Killing vector field $\x^a$. The vector field $z_i^a$ is also generated by this diffeomorphism, i.e.,  we have
\ba
\math{L}_\x z_i^a=0\,.
\ea
For the slice (cross section) $S$ is not the bifurcation surface $\math{B}$,  $z_i^a$ would be a finite vector field, i.e., contraction of any finite tensors is finite. Since $X_{a_1\cdots a_k}$ is smooth on the Killing horizon $\math{H}$, $X_{a_1\cdots a_k} z_1^a \cdots z_k^a$ would be finite on any cross section $S$.

When $S$ approach the bifurcation surface $\math{B}$, we have $\x^a\to 0$. However, note $s^a \x_a$ is finite,  $s^a$ must be divergent when $S\to\math{B}$. To show the divergence, we choose another two finite null vector field $k^a$ and $l^a$ on the cross-section $S$ near the bifurcation surface, which satisfies
\ba\begin{aligned}\label{kltoxs}
k^a l_a=&-1,\quad\quad k_ak^a=0,\quad\quad l_a l^a=0\,,\\
&k^a=C\x^a,\quad\quad l^a=C^{-1} s^a\,,
\end{aligned}\ea
in which $C$ is a scalar field on the cross section $S$. Since $k^a$ and $l^a$ are finite vectors on the bifurcation surface, we have $C\to \inf$ when $S\to\math{B}$. Denote $\bar{z}_i^a=\{k^a, l^a, w_i^a\}$. Since we assume that $X_{a_1\cdots a_k}$ is smooth near the bifurcation surface $\math{B}$, $X_{a_1\cdots a_k} \bar{z}_1^a \cdots \bar{z}_k^a$ will be finite on $\math{B}$. Since
\ba\begin{aligned}
\math{L}_{\x}X_{a_1\cdots a_k}=0\,,
\end{aligned}\ea
we have
\ba
\math{L}_\x(X_{a_1\cdots a_k} z_1^a \cdots z_k^a)=0\,,
\ea
which implies that $X_{a_1\cdots a_k} z_1^a \cdots z_k^a$ is invariance along the Killing vector $\x^a$. When the cross section $S$ is not the bifurcation surface $\math{B}$, $z^a$ would be a finite vector. Then, we have $X_{a_1\cdots a_k} z_1^a \cdots z_k^a$ is finite on whole future horizon $\math{H}$. From Eq. \eq{kltoxs}, we have
\ba\begin{aligned}\label{limitC}
X_{a_1\cdots a_k} \bar{z}_1^a \cdots \bar{z}_k^a=C^{m-n}X_{a_1\cdots a_k}\, z_1^a \cdots z_k^a\,,
\end{aligned}\ea
in which $m$ is the number of $\x^a$ and $n$ is the number of $s^a$. Since $X_{a_1\cdots a_k}$ is a smooth tensor near the bifurcation surface, $X_{a_1\cdots a_k} \bar{z}_1^a \cdots \bar{z}_k^a$ should be finite when $C$ approaches $\math{B}$. However, from Eq. \eq{limitC}, we can see that  when $m>n$, if $X_{a_1\cdots a_k}\, z_1^a \cdots z_k^a$ is finite, $X_{a_1\cdots a_k} \bar{z}_1^a \cdots \bar{z}_k^a$ would be divergent, which is in contradiction with the assumption that $X_{a_1\cdots a_k} \bar{z}_1^a$ is smooth near the bifurcation surface. Therefore, when $m>n$, we must have
\ba
X_{a_1\cdots a_k}\, z_1^a \cdots z_k^a=0\,.
\ea
As we desired to show.  $\Box$
\\

Since we assume that $E_R^{abcd}$ is smooth near horizon, $\grad_a E_R^{abcd}$ is also smooth near the horizon (including bifurcation surface). Using \textbf{Lemma 2}, Eq. \eq{first2} can reduce to
\ba\begin{aligned}\label{first3}
&\int_S \left[\bar{\d} \bm{Q}_\x-\x\.\bm{\Q}(\f,\d \f)\right]\\
=&\d\int_S\bm{\epsilon}_{aba_3\cdots a_n}\left[E_F^{ab}A^c\x_c-E_R^{abcd}\grad_{[c}\x_{d]}\right]-\int_S\x\.\left[\bm{\Theta}^\text{e.m}(\f,\d\bm{A})+\bm{\Theta}^\text{grav}(\f,\d g)\right] \,.
\end{aligned}\ea
For the first term of the left side in the above equation, considering that $\f_H=-A_a\x^a|_{\math{H}}$ is a constant on horizon, we have
\ba
\int_S\bm{\epsilon}_{aba_3\cdots a_n}E_F^{ab}A^c\x_c=-\F_H \int_\math{B} \bm{\epsilon}_{aba_3\cdots a_n} E_F^{ab}\,.
\ea
Using the on-shell condition $\grad_a E_F^{ab}=0$, it is not hard to get
\ba
-\int_\math{B} \bm{\epsilon}_{aba_3\cdots a_n} E_F^{ab}=-\int_\math{\inf} \bm{\epsilon}_{aba_3\cdots a_n} E_F^{ab}=Q\,.
\ea
Therefore, we have
\ba\begin{aligned}
\d\int_S\bm{\epsilon}_{aba_3\cdots a_n}E_F^{ab}A^c\x_c=\d(\F_H Q)\,.
\end{aligned}\ea
For the third term of Eq.\eq{first3}, we have
\ba\begin{aligned}
\int_S \x\.\bm{\Q}^\text{e.m.}(\f,\d \bm{A})&=2\int_S\x^c\bm{\epsilon}_{aca_3\cdots a_n}E_F^{ab}\d A_b\\
&=2\int_S \tilde{\bm{\epsilon}}\x^c\hat{\bm{\epsilon}}_{ac}E_F^{ab}\d A_b\,.
\end{aligned}\ea
Using $\hat{\bm{\epsilon}}=s\wedge\x=dv\wedge dr$, we can obtain
\ba\begin{aligned}
\int_S \x\.\bm{\Q}^\text{e.m.}(\f,\d \bm{A})&=2\int_S \tilde{\bm{\epsilon}}\x_aE_F^{ab}\d A_b=2\int_S \tilde{\bm{\epsilon}}\x_aE_F^{ab}s_b\x^c\d A_c\\
&=\int_S \tilde{\bm{\epsilon}}\hat{\bm{\epsilon}}_{ab}E_F^{ab}\x^c\d A_c=Q\d\F_H\,.
\end{aligned}\ea
In the following, we are going to evaluate the gravitational part. Since we choose the gauge such that $\x^a$ and $\math{H}$ is fixed in the variation, we can use the Gaussian null coordinates $\{v, r, \q^1,\cdots, \q^{n-2}\}$ to calculate these quantities. The line element of the spacetime in this coordinate can be expressed as\cite{Hollands:2006rj}
\ba\label{GNds2}
ds^2=2(dr-r \a dv-r\b_i d\q^i)dv+\g_{ij}d\q^i d\q^j\,,
\ea
in which $\a$, $\b_i$ and $\g_{ij}$ are the function of $r$ and $\q^i$. The horizon $\math{H}$ is determined by $r=0$. The Killing vector field generated the horizon is
\ba\begin{aligned}
\x^a=\left(\frac{\pd}{\pd v}\right)^a\,,\quad s^a=\left(\frac{\pd}{\pd r}\right)^a\,.
\end{aligned}\ea
Note that the gauge choice which fixes the Gaussian null coordinates is the same as the gauge choice to fix $\x^a$ and $\math{H}$. In this gauge choice, only $\a$, $\b_i$ and $\g_{ij}$ are dependent on the parameter $\l$. Based on the above coordinates, on the horizon $r=0$, we have
\ba
\grad_a\x_b=\k \hat{\bm{\epsilon}}_{ab}-\b_i\x_{[a} (d\q^i)_{b]} \,,
\ea
in which $\k=\a(0)$ is the surface gravity of the Killing horizon $\math{H}$. For the second term of Eq. \eq{first3}, we have
\ba\begin{aligned}
\int_S\bm{\epsilon}_{aba_3\cdots a_n}E_R^{abcd}\grad_{[c}\x_{d]}&=\int_S\tilde{\bm{\epsilon}}\hat{\bm{\epsilon}}_{ab}E_R^{abcd}\grad_{[c}\x_{d]}\\
&=\k\int_S\tilde{\bm{\epsilon}}\hat{\bm{\epsilon}}_{ab}\hat{\bm{\epsilon}}_{cd}E_R^{abcd}
+\int_S\tilde{\bm{\epsilon}}\b_i\hat{\bm{\epsilon}}_{ab}\x_c (d\q^i)_dE_R^{abcd}\,.
\end{aligned}\ea
Using \textbf{Lemma 2} and considering that $(d\q^i)^d$ is a tangent vector of $S$, the second term of the above expression vanishes. Thus we have
\ba\begin{aligned}
\int_S\bm{\epsilon}_{aba_3\cdots a_n}E_R^{abcd}\grad_{[c}\x_{d]}=\k\int_S\tilde{\bm{\epsilon}}\hat{\bm{\epsilon}}_{ab}\hat{\bm{\epsilon}}_{cd}E_R^{abcd}=-\frac{\k S}{2\p}\,.
\end{aligned}\ea
Therefore,  Eq. \eq{first3} reduces to
\ba\begin{aligned}
-\d\int_S\bm{\epsilon}_{aba_3\cdots a_n}E_R^{abcd}\grad_{[c}\x_{d]}=\frac{1}{2\p}\d(\k S)\,.
\end{aligned}\ea
For the last term of Eq. \eq{first3}, we have
\ba\begin{aligned}\label{lastexpression}
\int_S\x\.\bm{\Theta}^\text{grav}(\f,\d g)&=2\int_S\x^e\bm{\epsilon}_{ce\cdots a_n}\left(\grad_dE_R^{acbd}\d g_{ab}+E_R^{abcd}\grad_b \d g_{ad}\right)\\
&=-2\int_S\tilde{\bm{\epsilon}}\left(\x_c\grad_dE_R^{acbd}\d g_{ab}+ \x_cE_R^{abcd}\grad_b \d g_{ad}\right)\,.
\end{aligned}\ea
For the line element \eq{GNds2} in the Gaussian null coordinates, straightforward calculation gives $\d g_{ab}=\d \g_{ab}$, in which $\g_{ab}=\g_{ij}(d\q^i)_a(d\q^j)_b$. Based on \textbf{Lemma 2}, it is not hard to see that the first term of Eq. \eq{lastexpression} vanishes.
For the last term of Eq. \eq{lastexpression}, we have
\ba\begin{aligned}\label{sss43}
\x_cE_R^{abcd}\grad_b \d g_{ad}=-E_R^{\m\s\r r}\d g_{\m\r; \s}\,.
\end{aligned}\ea
Using the line element \eq{GNds2}, \textbf{Lemma 2} can be presented as: if $E_R^{\m\s\r\t}$ is not zero on horizon, the number of ``$v$'' in the index must be larger than number of ``$r$''. Therefore, the nonvanishing contributions in Eq. \eq{sss43} from $\d g_{\m\r; \s}$ only comes from the term in which the number of ``$v$'' larger that number of ``$r$''. Using the line element, it is not hard to find that the only nonvanishing component of $\d g_{\m\r; \s}$ is
\ba
\d g_{vv; r}=-2\d\a=-2\d \k\,.
\ea
Therefore, we have
\ba\begin{aligned}\label{sss44}
\int_S\x\.\bm{\Theta}^\text{grav}(\f,\d g)&=-4\d \k\int_S \tilde{\bm{\epsilon}} E_R^{vrvr}=\frac{1}{2\p}S\d\k\,.
\end{aligned}\ea
Summing the above results, we have
\ba\begin{aligned}
\d M=\frac{\k}{2\p} \d S+\W_H^{(\m)}\d J_{(\m)}+\F_H \d Q\,.
\end{aligned}\ea
Therefore, we derived the first law of black holes in the gravitational electromagnetic system and the expression is the same as the Einstein-Maxwell theory.

\section{Conclusion}\label{sec4}

The first law of black hole in a diffeomorphism covariant theory is generally derived by Iyer and Wald \cite{Iyer:1994ys}. However, their results bases on the requirement that all dynamical fields be smooth near the future Killing horizon as well as the bifurcation surface, and consequently the ``potential-charge'' term does not appear in their result for the gravitational electromagnetic system. In this paper, we extended their discussion into the gravitational electromagnetic system without the requirement that the vector potential $\bm{A}$ is smooth near the horizon since it is not a real physical quantity.

Firstly, we calculated the Noether charge and variational identity of the gravitational electromagnetic theory with the Lagrangian $\math{L}(g_{ab}, R_{abcd}, F_{ab})$. Then, using these results, we derived the thermodynamic first law of the asymptotically flat stationary axisymmetric black holes. In contrast to the earlier discussion by Iyer and Wald, we only require that the electromagnetic strength $F_{ab}$ and metric $g_{ab}$ be smooth near the future horizon (including the bifurcation surface), without making any constraints for the vector potential $\bm{A}$. Under the above conditions, we obtain the first law of thermodynamics for the mass, angular momentum, and charge variation of the black hole. The result is the same as the expression of the first law in Einstein-Maxwell theory. Our investigation shows that the first law of black hole thermodynamics is also universal in an effective theory that takes into account the quantum corrections in Einstein-Maxwell theory, and Wald entropy is still the best choice to describe the entropy of a steady black hole.

\noindent
\textbf{Acknowledgments}
We acknowledge financial supports from the National Natural Science Foundation of China (Grants No. 11775022, 11873044 and 12005080).


\begin{thebibliography}{100}
\bibitem{A2}
  J.~D.~Bekenstein,``Black holes and the second law,''  Lett.\ Nuovo Cim.\  {\bf 4}, 737 (1972).
\bibitem{A31}
J. M. Bardeen, B. Carter and S. W. Hawking, ``The Four laws of
black hole mechanics,'' Commun. Math. Phys. {\bf31}, 161 (1973).
\bibitem{A3}
  J.~D.~Bekenstein, ``Generalized second law of thermodynamics in black hole physics,'' Phys.\ Rev.\ D {\bf 9}, 3292 (1974).
\bibitem{A4}
  S.~W.~Hawking, ``Particle Creation by Black Holes,'' Commun.\ Math.\ Phys.\  {\bf 43}, 199 (1975).
\bibitem{Sudarsky:1992ty}
D.~Sudarsky and R.~M.~Wald, ``Extrema of mass, stationarity, and staticity, and solutions to the Einstein Yang-Mills equations,''
Phys. Rev. D \textbf{46}, 1453-1474 (1992).

\bibitem{Iyer:1994ys}
V.~Iyer and R.~M.~Wald,``Some properties of Noether charge and a proposal for dynamical black hole entropy,''
Phys. Rev. D \textbf{50}, 846-864 (1994).

\bibitem{Gao:2003ys}
S.~Gao, ``The First law of black hole mechanics in Einstein-Maxwell and Einstein-Yang-Mills theories,''
Phys. Rev. D \textbf{68}, 044016 (2003).

\bibitem{A13}
 B.~Zwiebach, ``Curvature Squared Terms and String Theories,'' Phys.\ Lett.\  {\bf 156B}, 315 (1985).
\bibitem{A14}
 D.~J.~Gross and E.~Witten, ``Superstring Modifications of Einstein's Equations,'' Nucl.\ Phys.\ B {\bf 277}, 1 (1986).
\bibitem{A15}
 A.~Sen, ``Black Hole Entropy Function, Attractors and Precision Counting of Microstates,'' Gen.\ Rel.\ Grav.\  {\bf 40}, 2249 (2008).
\bibitem{A16}
A.~Dabholkar and S.~Nampuri, ``Quantum black holes,'' Lect.\ Notes Phys.\  {\bf 851}, 165 (2012).
\bibitem{Wald:1999wa}
R.~M.~Wald and A.~Zoupas, ``A General definition of 'conserved quantities' in general relativity and other theories of gravity,'' Phys. Rev. D \textbf{61}, 084027 (2000).
\bibitem{Hollands:2006rj}
S.~Hollands, A.~Ishibashi and R.~M.~Wald, ``A Higher dimensional stationary rotating black hole must be axisymmetric,''
Commun. Math. Phys. \textbf{271}, 699-722 (2007).
\end{thebibliography}
\end{document}